\documentclass[aps,prb,graphicx,preprint]{revtex4-2}

\usepackage{hyperref}
\usepackage{amsmath}
\usepackage{array}
\usepackage{dcolumn}
\usepackage{graphicx}
\usepackage{grffile}
\usepackage{siunitx}
\usepackage{mw}
\usepackage{bm}
\usepackage{xcolor}
\usepackage{multirow}
\usepackage{float}
\usepackage{caption}
\usepackage{subcaption}
\begin{document}

\title{First principles residual resistivity using locally self-consistent multiple scattering method}

\author{Vishnu Raghuraman}
\affiliation{Department of Physics, Carnegie Mellon University, Pittsburgh, PA 15213}

\author{Markus Eisenbach}
\affiliation{Oak Ridge National Laboratory, Oak Ridge, TN 37831}

\author{Michael Widom}
\affiliation{Department of Physics, Carnegie Mellon University, Pittsburgh, PA 15213}

\author{Yang Wang}
\affiliation{Pittsburgh Supercomputing Center, Carnegie Mellon University, Pittsburgh, PA 15213}

\begin{abstract}
The locally self-consistent multiple scattering (LSMS) method can perform efficient first-principles calculations of systems with large number of atoms. In this work, we combine the Kubo-Greenwood equation with LSMS, enabling us to calculate first-principles residual resistivity of large systems. This has been implemented in the open-source code \textit{lsms}. 
We apply this method to selected pure elements and binary random alloys.
The results compare well with experiment, and with values obtained from a first-principles effective medium technique (KKR-CPA). We discuss future applications of this method to complex systems where other methods are not applicable.
\end{abstract}
\maketitle
\section{Introduction}
Disordered systems show interesting physical and chemical properties \cite{miracle-senkov,metallic-glass,quasicrystals}. Multi-principal element alloys with high chemical disorder have shown high strength and ductility over large temperature ranges \cite{mpea-strength-ductility-1,mpea-strength-ductility-2,mpea-strength-ductility-3}. Bulk metallic glasses exhibit ultra-high strength, high elasticity, high fracture toughness, high wear resistance and other useful properties \cite{metallic-glass-properties-1,metallic-glass-properties-2}. Quasicrystals possess high thermal and electrical resistivity, low adhesion, and have been used as coating for non-stick cookware \cite{quasicrystal-properties}. First-principles density functional theory allows us to compute phase behavior, band structure, mechanical and functional properties for these systems \cite{hea-dft-1,hea-dft-2,hea-dft-3}. However, the computational cost of DFT calculations grows as the cube of the system size, making the study of large systems impractical. It is possible to use an effective medium method, such as the coherent potential approximation (CPA) \cite{soven} to model the system using few atoms. While this works well for chemical species disorder, it is difficult to construct an effective medium for atomic displacements, or other forms of disorder which break crystallinity. Alternatively, we can perform classical MD using potentials fitted against DFT data. This approach is significantly faster in comparison to DFT, however, obtaining these potentials is a highly difficult task, especially for complex systems, and classical MD yields no information concerning electronic structure.

The locally self-consistent multiple scattering (LSMS) method \cite{LSMS-method}, based on the Korringa-Kohn-Rostoker (KKR) Green's function approach to DFT \cite{korringa,kohnrostoker}, offers an efficient solution to this problem. As in the KKR method, LSMS uses multiple scattering theory to obtain the Green's function of the system. However, electron scattering between widely separated atoms is ignored. The cut-off distance for non-zero scattering is represented by the local interaction zone (LIZ) radius.
This approximation speeds up the calculation significantly. LSMS scales linearly with the system size, making it a practical computational tool for disorder studies. 

The KKR Green's function method can be combined with the Kubo-Greenwood linear response formula\cite{kubo,greenwood} to obtain first-principles electrical conductivity. The Kubo-Greenwood equation depends on products of Green's functions, which are readily available in KKR. For a random system, this product must be averaged over the ensemble. Butler \cite{butler} showed that the ensemble average can be treated using the CPA medium. The KKR-CPA conductivity method has since been applied successfully to several systems \cite{swihart,saimu,cpa-conductivity-paper}. Alternatively, we can represent the ensemble average using a single, carefully constructed large structure. The Green's function for this system can be calculated using LSMS and inserted into the Kubo-Greenwood equation, which produces the electrical conductivity. This linear scaling nature of LSMS enables us to apply this approach to systems with tens of thousands of atoms, allowing us to calculate conductivity for structures with intricate features like stacking faults, dislocations and quasicrystalline order. We have implemented this in the open source high-performance software package $lsms$ \cite{lsms}. The resulting conductivity depends on the LIZ radius, and convergence with LIZ radius will be a major point of discussion in this work.

The next section provides some theoretical background on the LSMS method and the Kubo-Greenwood equation. We then introduce our LSMS resistivity approach and provide a heuristic derivation for the conductivity as a function of the LIZ radius.
We test our method by applying it to selected pure elements with a variety of structural and electronic characteristics (Ag, Al, Li and V) and binary alloys (Al$_x$V$_{1-x}$ and Fe-9 wt\%Si). We compare the results with experimental data and computational values obtained from KKR-CPA. Finally, we conclude by discussing areas of improvements for the method and potential future applications. 

\section{Methods}

\subsection{KKR and LSMS}
The Korringa-Kohn-Rostoker (KKR) approach to density functional theory solves for the Green's function of the Kohn-Sham equation \cite{dft-1,dft-2}. The charge density can be obtained from the Green's function using
\begin{equation}
    \rho(\bm{r}) = -\frac{1}{\pi}\int_{-\infty}^{\epsilon_{F}} d\epsilon\;\Im\left[\mathrm{Tr}(G(\bm{r}, \bm{r}, \epsilon))\right],
\end{equation}
which is then used to calculate a new Hamiltonian. Using multiple scattering theory, we express the Green's function in the vicinity of atomic site $n$ as \cite{faulkner_stocks_1980,faulkner_stocks_wang}
\begin{equation}
    G(\bm{r}_n, \bm{r}_n, \epsilon) =  \sum_{LL^{\prime}} Z^{n}_{L}(\bm{r}_n, \epsilon)\tau^{nn}_{LL^{\prime}}(\epsilon)Z^{n}_{L^{\prime}}(\bm{r}_n) - \sum_{L} Z^{n}_{L}(\bm{r}_n,\epsilon)J^{n*}_{L}(\bm{r}_{n}, \epsilon),
\end{equation}
where $Z^{n}$ and $J^{n}$ are the regular and irregular solutions to the single-atom Schr\"{o}dinger equation for the atom at site $n$. $L$ and $L'$ are angular momentum indices ({\em e.g.} $L\equiv(lm)$). The multiple scattering path matrix $\tau$ can be expressed in terms of the single-site scattering $t$-matrix and the free electron propagator $g$ as \cite{tau-def}
\begin{align}
    \label{eq:tau-inverse}
    \underline{\tau}^{mn} &= \left(\left[\underline{T}^{-1} - \underline{g}\right]^{-1}\right)^{mn}, \\
    \underline{T}^{mn} &= \underline{t}^{m}\delta^{mn}.
\end{align}
Here $m$ and $n$ refer to atomic sites. Formally, $\tau$ obeys a Dyson expansion
\begin{align}
\tau^{mn} &= t^m\delta^{mn} + t^m g^{mn} t^n + \sum_{k\ne m,n} t^m g^{mk} t^k g^{kn} t^n + \cdots \\
&= t^m\delta^{mn} + t^m g^{mk} \tau^{kn}.
\label{eq:Dyson}
\end{align}

Calculating the matrix inverse in Equation \ref{eq:tau-inverse} is the most computationally intensive step in the KKR method. This operation scales as the cube of the number of atoms in the system.
\begin{figure}
    \centering
    \includegraphics[width=0.5\linewidth]{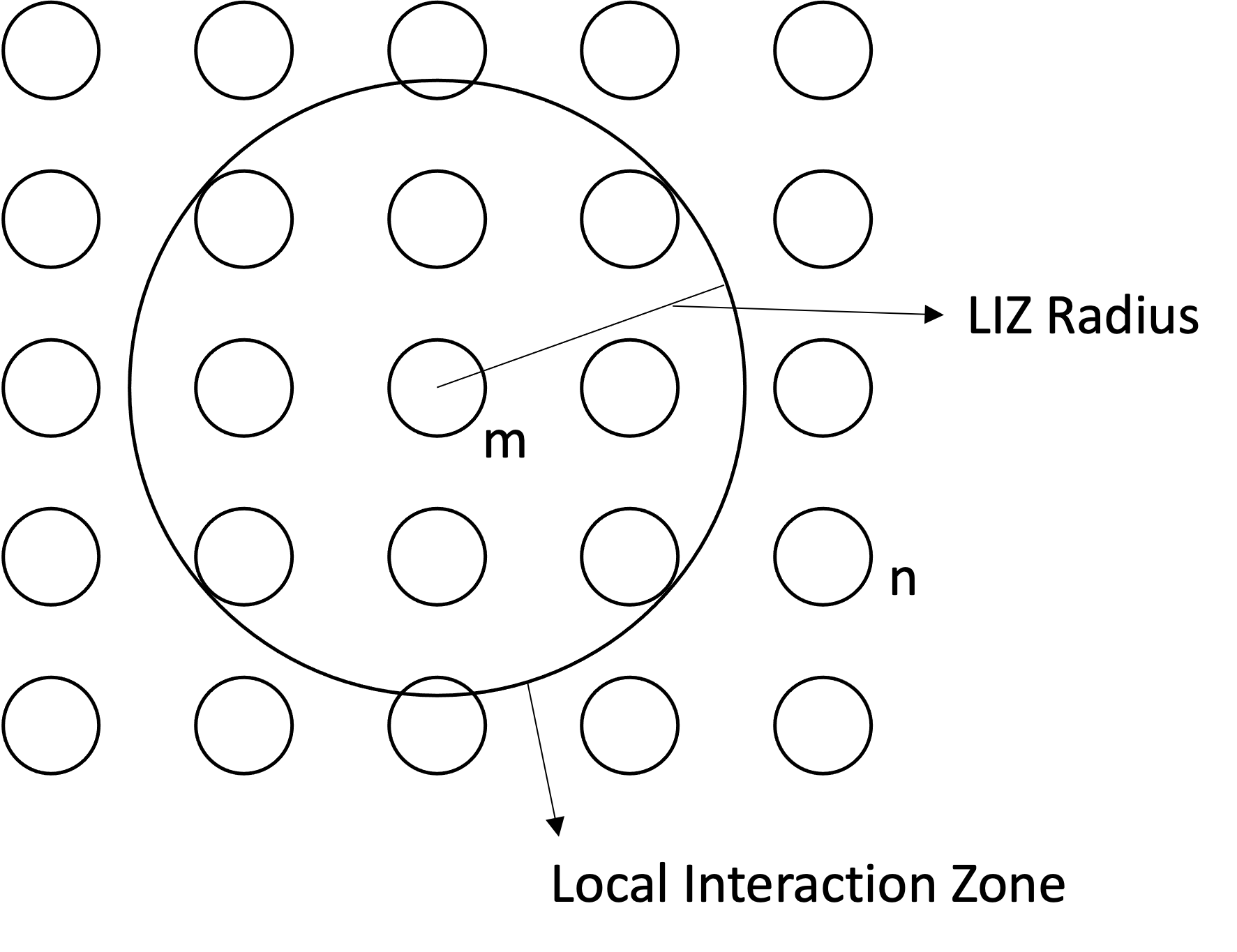}
    \caption{Schematic depiction the local interaction zone (LIZ) for an atom $m$. Scattering between $m$ and any atom $n$ beyond the LIZ is ignored ($\tau^{mn} = 0$)}
    \label{fig:liz-schematic}
\end{figure}
In LSMS, we define a local interaction zone (LIZ) for each atom, beyond which scattering is neglected (Figure \ref{fig:liz-schematic}). This results in a much smaller $\tau$ matrix, and the inverse scales linearly with the system size. It is important to choose an appropriate value of LIZ radius. A large LIZ radius slows down the calculation, while a small LIZ radius produces inaccurate results. For a given system, it is important to test multiple LIZ radii to ensure convergence.

\subsection{Conductivity}
Electrical conductivity obeys the Kubo-Greenwood equation \cite{butler}
\begin{align}
     \label{eq:sigma-tilde-consolidation}
    \sigma_{\mu\nu} = \frac{1}{4}\lim_{\delta \rightarrow 0} &\left[\tilde{\sigma}_{\mu\nu}(\epsilon_F + i\delta,\epsilon_{F} + i\delta) - \tilde{\sigma}_{\mu\nu}(\epsilon_F + i\delta, \epsilon_F - i\delta)\right. \\
    &-\left.\tilde{\sigma}_{\mu\nu}(\epsilon_F - i\delta, \epsilon_F + i\delta) + \tilde{\sigma}_{\mu\nu}(\epsilon_F - i\delta, \epsilon_F - i\delta)\right], \nonumber\\
    \tilde{\sigma}_{\mu\nu}(z_1, z_2) = -&\frac{\hbar}{\pi N\Omega} \mathrm{Tr}\left<j_{\mu}G(z_1)j_{\nu}G(z_2)\right>,
\end{align}
where $\mu$ and $\nu$ refer to Cartesian directions, $j_\mu$ and $j_{\nu}$ are current operators, $N$ is the number of atoms, and $\Omega$ is the atomic volume. The angular brackets represent an ensemble average over different random configurations. We express this equation in terms of multiple scattering matrices as \cite{butler} 
\begin{equation}
    \tilde{\sigma}_{\mu\nu}(z_1, z_2) = -\frac{4m_e^2}{\pi N \Omega\hbar^3}\sum_{mn}\sum_{L_1L_2L_3L_4} \left<J^{m\mu}_{L_4L_1}(z_2, z_1)\tau^{mn}_{L_1L_2}(z_1)J^{n\nu}_{L_2L_3}(z_1, z_2)\tau^{nm}_{L_3L_4}(z_2)\right>.
\end{equation}
where $J$ is the matrix element of the current operator. Since the sum over $m$ generates a volume average, we replace the ensemble average with a single, sufficiently large and representative, random structure. Additionally, for each atom $m$, we only consider atoms $n$ within the LIZ of $m$. Within the LSMS formalism,
\begin{equation}
    \tilde{\sigma}_{\mu\nu}(z_1, z_2) = -\frac{4m_e^2}{\pi N\Omega\hbar^3}\sum_{m}^{N}\sum_{n}^{\mathrm{LIZ}_{m}} \sum_{L_1L_2L_3L_4} J^{m\mu}_{L_4L_1}(z_2, z_1)\tau^{mn}_{L_1L_2}(z_1)J^{n\nu}_{L_2L_3}(z_1, z_2)\tau^{nm}_{L_3L_4}(z_2).
    \label{eq:lsms-conductivity-formula}
\end{equation}
Combining equations \ref{eq:lsms-conductivity-formula} and \ref{eq:sigma-tilde-consolidation} yields the LSMS electrical conductivity tensor for a given LIZ size, which is then inverted to obtain the resistivity. 

\subsection{Dependence on LIZ radius}
Because our LIZ sizes are limited by available computing resources, we wish to model the convergence of conductivity with increasing LIZ size. Here we derive this dependence heuristically. 
Consider a perfectly periodic crystal for which the electron mean free path is infinite. Assuming weak scattering, and truncating the Dyson expansion Eq.~(\ref{eq:Dyson}) at first order in $g$, we find the $L=0$ component of the $\tau$-matrix varies asymptotically as
\begin{equation}
    \tau_{L_1L_2}(r, z) \approx \beta_{L_1L_2}(z) \frac{e^{i\kappa r}}{r}
\end{equation}
where $\kappa = \sqrt{2m_e z/\hbar^2}$. 

Replacing the sums in Eq.~(\ref{eq:lsms-conductivity-formula}) with integrals, the conductivity components $\tilde{\sigma}(z_1,z_2)$ can be expressed as
\begin{equation}
    \tilde{\sigma}_{\mu\nu}(z_1,z_2) \approx A(z_1,z_2)\int_{r_c}^{r_{\rm{LIZ}}} dr\;r^2\left(\frac{e^{i\kappa_1 r}}{r}\right)\left(\frac{e^{i\kappa_2 r}}{r}\right)
\end{equation}
where $\kappa_{1,2} = \sqrt{2m_ez_{1,2}/\hbar^2}$, $r_c$ is the nearest neighbor separation and $r_{\rm{LIZ}}$ is the LIZ radius. We can integrate this to obtain the four terms needed for conductivity
\begin{align}
    \tilde{\sigma}_{\mu\nu}(r_{\rm{LIZ}}, \epsilon_{+}, \epsilon_{+}) &\approx \frac{A_{++}}{\kappa_{+}}e^{i\kappa_{+}(r_{\rm{LIZ}} + r_{c})}\sin(\kappa_{+}(r_{\rm{LIZ}} - r_{c})) \\
    \tilde{\sigma}_{\mu\nu}(r_{\rm{LIZ}}, \epsilon_{+}, \epsilon_{-}) &\approx A_{+-}\left[r_{\rm{LIZ}} - r_{c}\right] \\
    \tilde{\sigma}_{\mu\nu}(r_{\rm{LIZ}}, \epsilon_{-}, \epsilon_{+}) &\approx A_{-+}\left[r_{\rm{LIZ}} - r_c\right] \\
    \tilde{\sigma}_{\mu\nu}(r_{\rm{LIZ}}, \epsilon_{-}, \epsilon_{-}) &\approx \frac{A_{--}}{\kappa_{+}}e^{-i\kappa_{+}(r_{\rm{LIZ}} + r_c)}\sin(\kappa_{+}(r_{\rm{LIZ}} - r_c))
\end{align}
where $\epsilon_{+-} = \epsilon_{F} \pm i\delta$, $\kappa_{+} = \sqrt{2m_e \epsilon_{+}/\hbar^2}$ and we have taken the limit $\delta\to 0$. The conductivity is given by
\begin{equation}
    \sigma_{\mu\nu} \approx -\frac{\left[A_{+-} + A_{-+}\right]}{4}(r_{\rm{LIZ}} - r_c) + \frac{\sin{(\kappa_{+}(r_{\rm{LIZ}} - r_{c}))}}{4k_{+}}\left[A_{++}e^{i\kappa_{+}(r_{\rm{LIZ}} - r_c)} + A_{--}e^{i\kappa_{-}(r_{\rm{LIZ}} - r_c)}\right]
\end{equation}
The conductivity shows a combination of linear and oscillatory behavior. The oscillations occur at frequency $2k_F$ with $k_F$ the Fermi wavenumber.
At large $r_{\rm{LIZ}}$ the linear term dominates, the conductivity diverges and the resistivity vanishes, as expected for perfectly crystalline systems. In particular, the resistivity vanishes linearly as a function of the inverse LIZ radius.

In order for the conductivity to converge, the $\tau$-matrix must decay faster than $1/r$. To model systems with finite conductivity, consider the following ansatz
\begin{equation}
    \tau_{L_1L_2}(z) \approx \beta_{L_1L_2}(z)\frac{e^{i\kappa r}}{r}e^{-\alpha r}
\end{equation}
where $\alpha$ is a decay parameter determined by the inverse of the mean free path. For weak disorder, $\alpha \rightarrow 0$ and we recover the linear conductivity expression. For disordered systems, $\alpha>0$ and the conductivity component $\tilde{\sigma}$ becomes
\begin{equation}
    \tilde{\sigma}_{\mu\nu}(z_1,z_2) \approx A(z_1,z_2)\int_{r_c}^{r_{\rm{LIZ}}} dr\;r^2\left(\frac{e^{i\kappa_1r}}{r}\right)\left(\frac{e^{i\kappa_2r}}{r}\right)e^{-2\alpha r}
\end{equation}
resulting in
\begin{align}
    \tilde{\sigma}_{\mu\nu}(\epsilon_{+}, \epsilon_{+}) &\approx A_{++}\left[\frac{e^{(2i\kappa_{+} - 2\alpha)r_{\rm{LIZ}}} - e^{(2i\kappa_{+} - 2\alpha)r_c}}{2i\kappa_{+} - 2\alpha}\right] \\
    \tilde{\sigma}_{\mu\nu}(\epsilon_{+}, \epsilon_{-}) &\approx A_{+-}\left[\frac{e^{-2\alpha r_c} - e^{-2\alpha r_{\rm{LIZ}}}}{2\alpha}\right] \\
    \tilde{\sigma}_{\mu\nu}(\epsilon_{-}, \epsilon_{+}) &\approx A_{-+}\left[\frac{e^{-2\alpha r_c} - e^{-2\alpha r_{\rm{LIZ}}}}{2\alpha}\right] \\
    \tilde{\sigma}_{\mu\nu}(\epsilon_{-}, \epsilon_{-}) &\approx A_{--}\left[\frac{e^{-(2i\kappa_{+} + 2\alpha)r_{\rm{LIZ}}} - e^{-(2i\kappa_{+} + 2\alpha)r_c}}{-2i\kappa_{+} - 2\alpha}\right].
\end{align}
Note the combination of oscillating and exponentially decaying terms. The conductivity approaches a finite limit as
\begin{equation}
    \sigma_{\mu\nu} \approx a_0 e^{-2\alpha r_{\rm{LIZ}}} + a_1
    \label{eq:sigma-alpha}
\end{equation}
where
\begin{align}
    a_0 &= \frac{1}{4}\left[\frac{A_{++}}{2ik_{+} - 2\alpha}e^{2ik_{+}r_{\rm{LIZ}}} - \frac{A_{--}}{2ik_{+} + 2\alpha}e^{-2i\kappa_{+}r_{\rm{LIZ}}} + \frac{(A_{+-} + A_{-+})}{2\alpha}\right] \\
    a_1 &= -\frac{1}{4}\left[\frac{A_{++}}{2i\kappa_{+} - 2\alpha}e^{2i\kappa_{+}r_c} - \frac{A_{--}}{2i\kappa_{+} + 2\alpha}e^{-2i\kappa_{+}r_c} + \frac{(A_{+-} + A_{-+})}{2\alpha}\right]e^{-2\alpha r_c}
\end{align}
The resistivity is now finite, and its dependence on the inverse LIZ size is {\em non}-linear.

To recover the limit of weak disorder, with the linear variation of resistivity, take $\alpha r_{\rm LIZ} \ll 1$. In this case we can write
\begin{equation}
    a_0 \approx \frac{A_{+-} + A_{-+}}{8\alpha},\;
    a_1 \approx -\frac{A_{+-} + A_{-+}}{8\alpha}e^{-2\alpha r_c},\;
    a_0 \approx -a_1 e^{2\alpha r_{c}}.
\end{equation}
Inserting this in the conductivity expression Eq.~(\ref{eq:sigma-alpha}) gives us
\begin{align}
    \sigma_{\mu\nu} &\approx a_1 \left(1 - e^{-2\alpha(r_{\rm{LIZ}} - r_c)}\right) \\
    &\approx \frac{2a_1\alpha (r_{\rm{LIZ}} - r_c)}{1 + 2\alpha(r_{\rm{LIZ}} - r_c)}.
\end{align}
To see the linearity of resistivity, invert $\sigma$ to obtain
\begin{align}
    \rho_{\mu\nu} &\approx \frac{1 + 2\alpha(r_{\rm{LIZ}} - r_c)}{2\alpha a_1 (r_{\rm{LIZ}} - r_c)} \\
    &\approx \frac{1}{2\alpha a_1}k_{\rm{LIZ}} + \frac{1}{a_1}\;,\;k_{\rm{LIZ}} = \frac{1}{r_{\rm{LIZ}} - r_c}
    \label{eq:linear-fit}
\end{align}
In the small $\alpha$ regime, the resistivity varies linearly with the inverse of the LIZ radius, with the intercept being the converged value.

For $r_{\rm{LIZ}} \rightarrow \infty$, the above expression~\ref{eq:linear-fit} breaks down as $\alpha r_{\rm{LIZ}}$ increases. But as $\alpha \rightarrow 0$, the LIZ radius range over which we see linearity will increase. As a result, using the linear expression in lieu of the full non-linear form might result in smaller errors. We will explore this idea further in the next section when we discuss the resistivity of binary random alloys.

\section{Computational Details}
We apply the LSMS resistivity method (as implemented in the open source code $lsms$) to a 20x20x20 supercell (16,000 atoms) for BCC structures and a 16x16x16 supercell (16,384 atoms) for FCC structures. A suitable starting potential is needed for resistivity calculations. For the pure elements and Al$_x$V$_{1-x}$ binaries, we perform self-consistent LSMS calculations to obtain a converged potential. For FeSi, we obtain converged atomic potentials for Fe and Si from a KKR-CPA calculation. We employ the von Barth-Hedin local density approximation \cite{dft-1,vbh} for the exchange-correlation functional. KKR-CPA calculations were performed using open source code MuST \cite{must}. For FeSi, we perform spin-polarized calculations. Computationally intensive sections of the calculations are GPU accelerated. All LSMS calculations are performed on the Frontier supercomputer at ORNL. The calculations are incredibly efficient and scalable - using 1 GPU per atom, we are able to calculate resistivity for a 16,000 atom structure with approximately a 1000 atom LIZ in under 10 minutes.
\section{Results}
\subsection{Pure Elements}
\begin{figure}
    \centering
    \begin{subfigure}[b]{0.49\textwidth}
       \includegraphics[width=\linewidth]{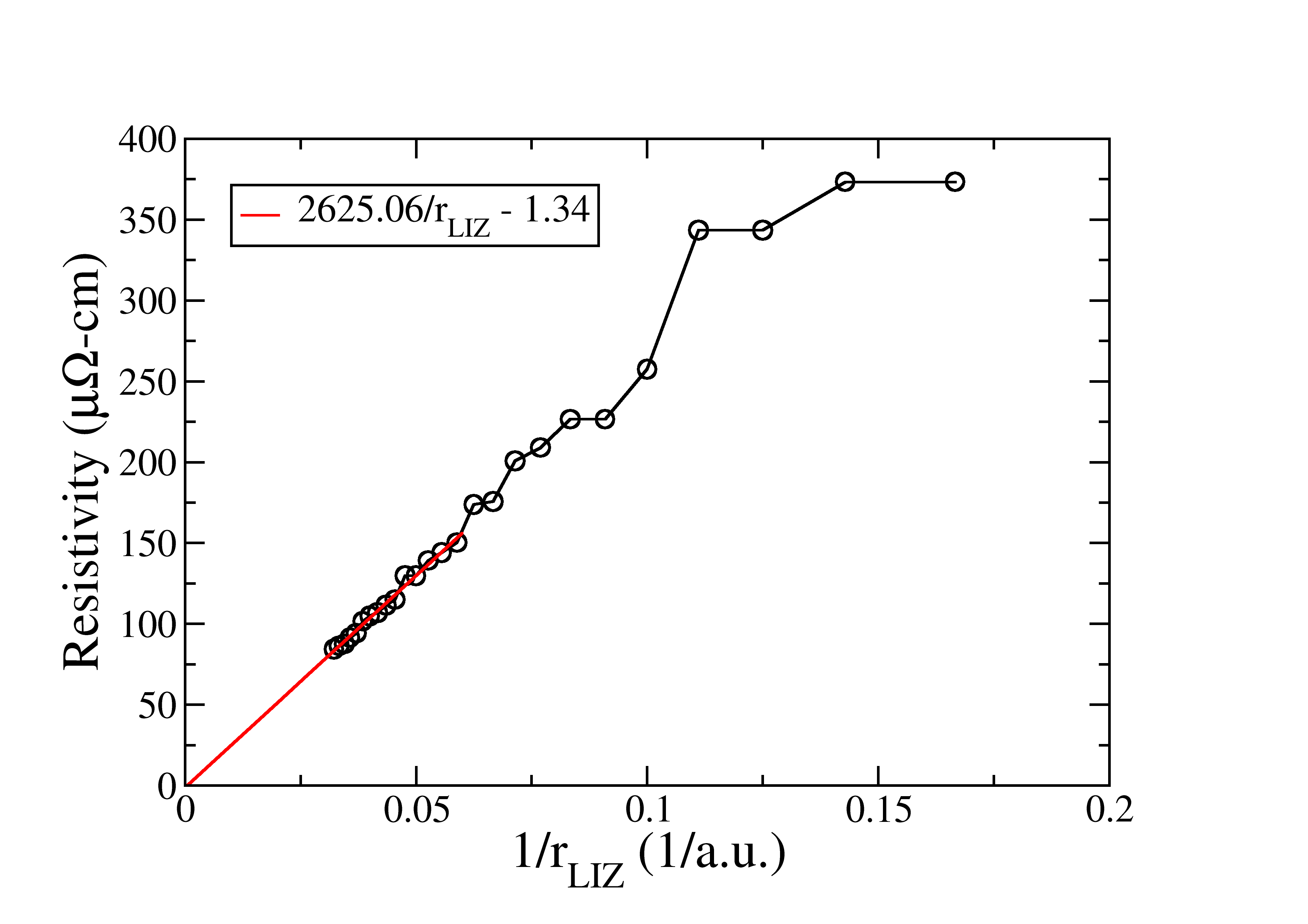}
       \caption{Ag}
       \label{fig:Ag-resistivity}
    \end{subfigure}
    \begin{subfigure}[b]{0.49\textwidth}
       \includegraphics[width=\linewidth]{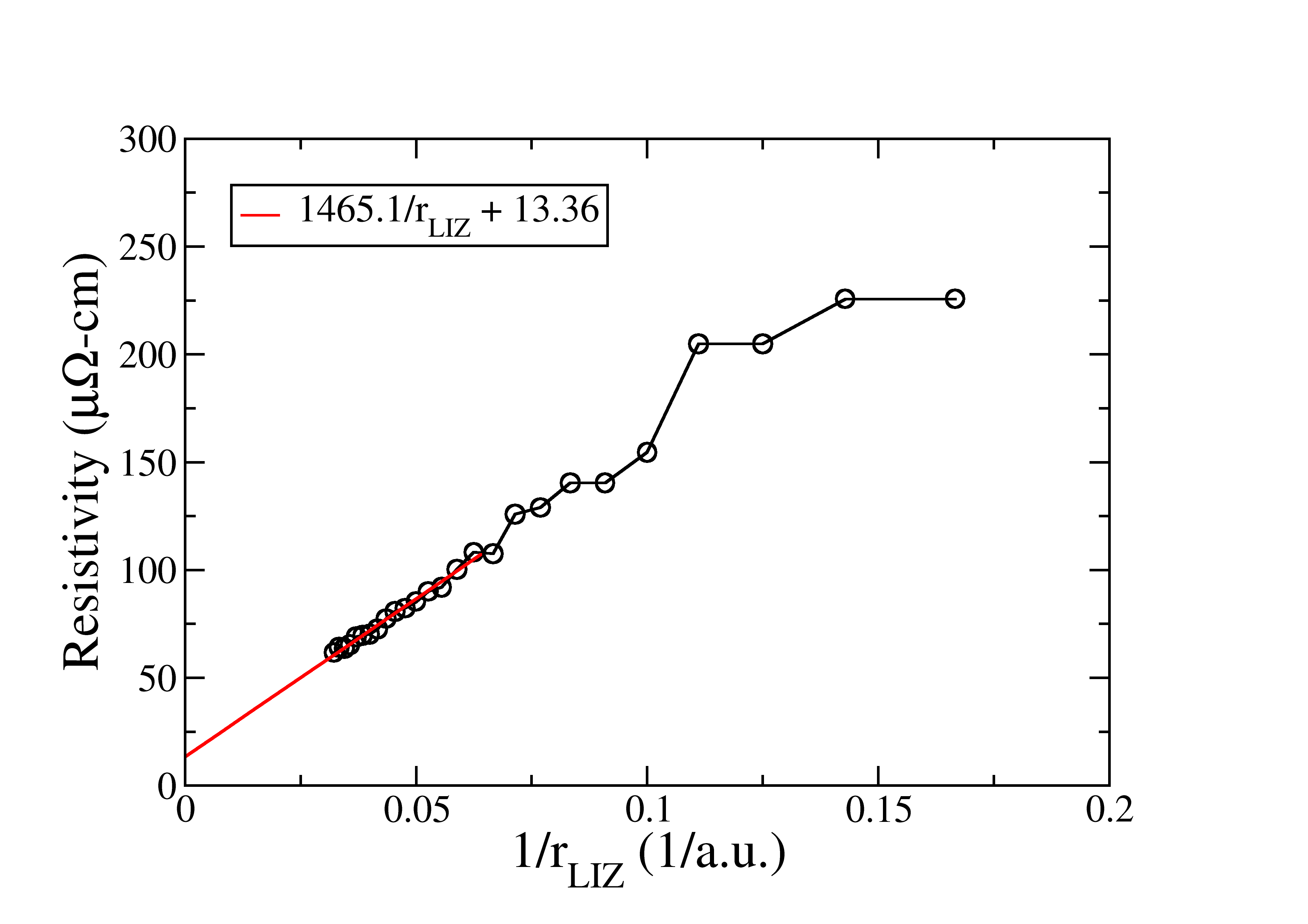}
       \caption{Al}
       \label{fig:Al-resistivity}
    \end{subfigure}
    \begin{subfigure}[b]{0.49\linewidth}
       \includegraphics[width=\linewidth]{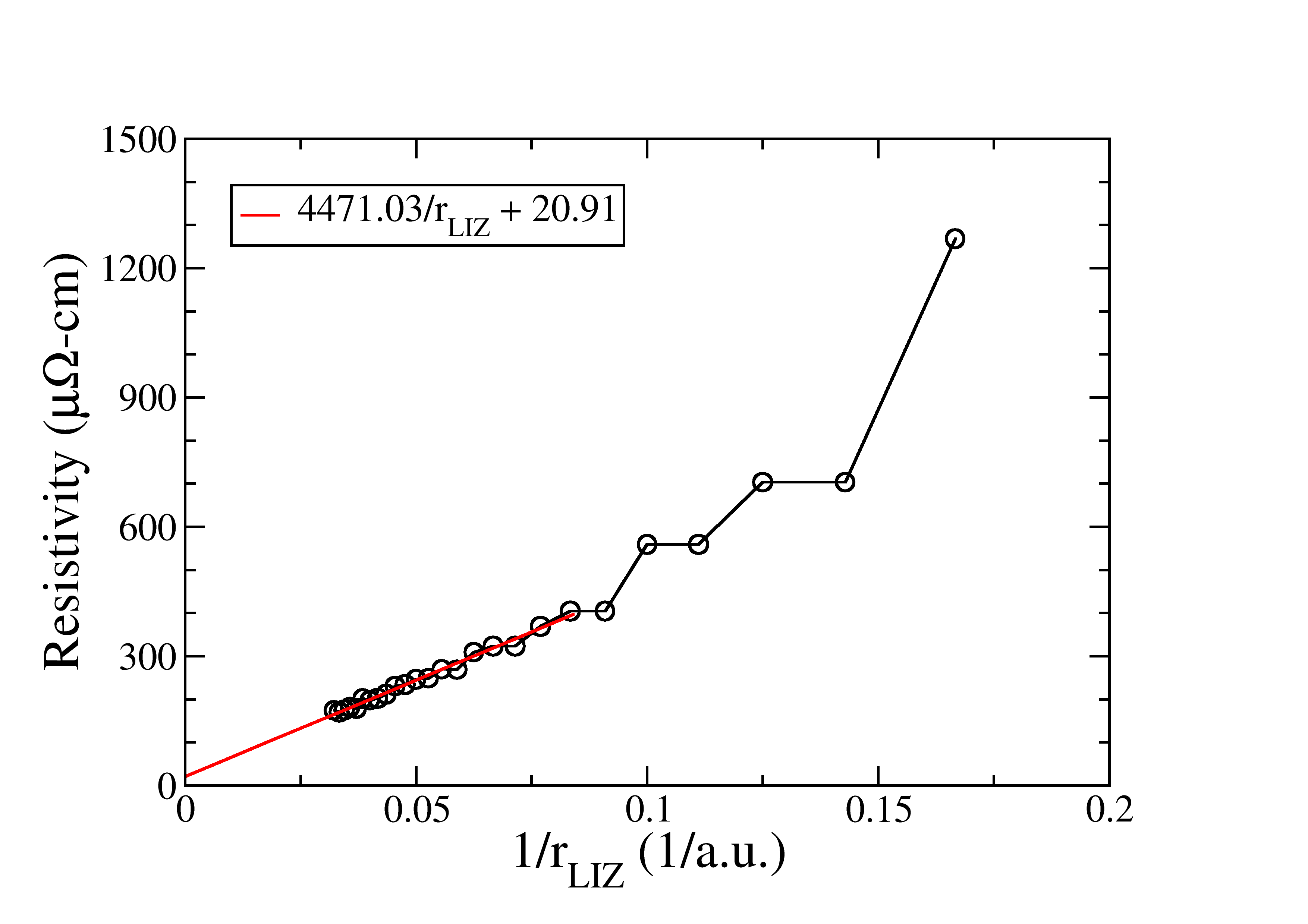}
       \caption{Li}
       \label{fig:Li-resistivity}
    \end{subfigure}
    \begin{subfigure}[b]{0.49\linewidth}
       \includegraphics[width=\linewidth]{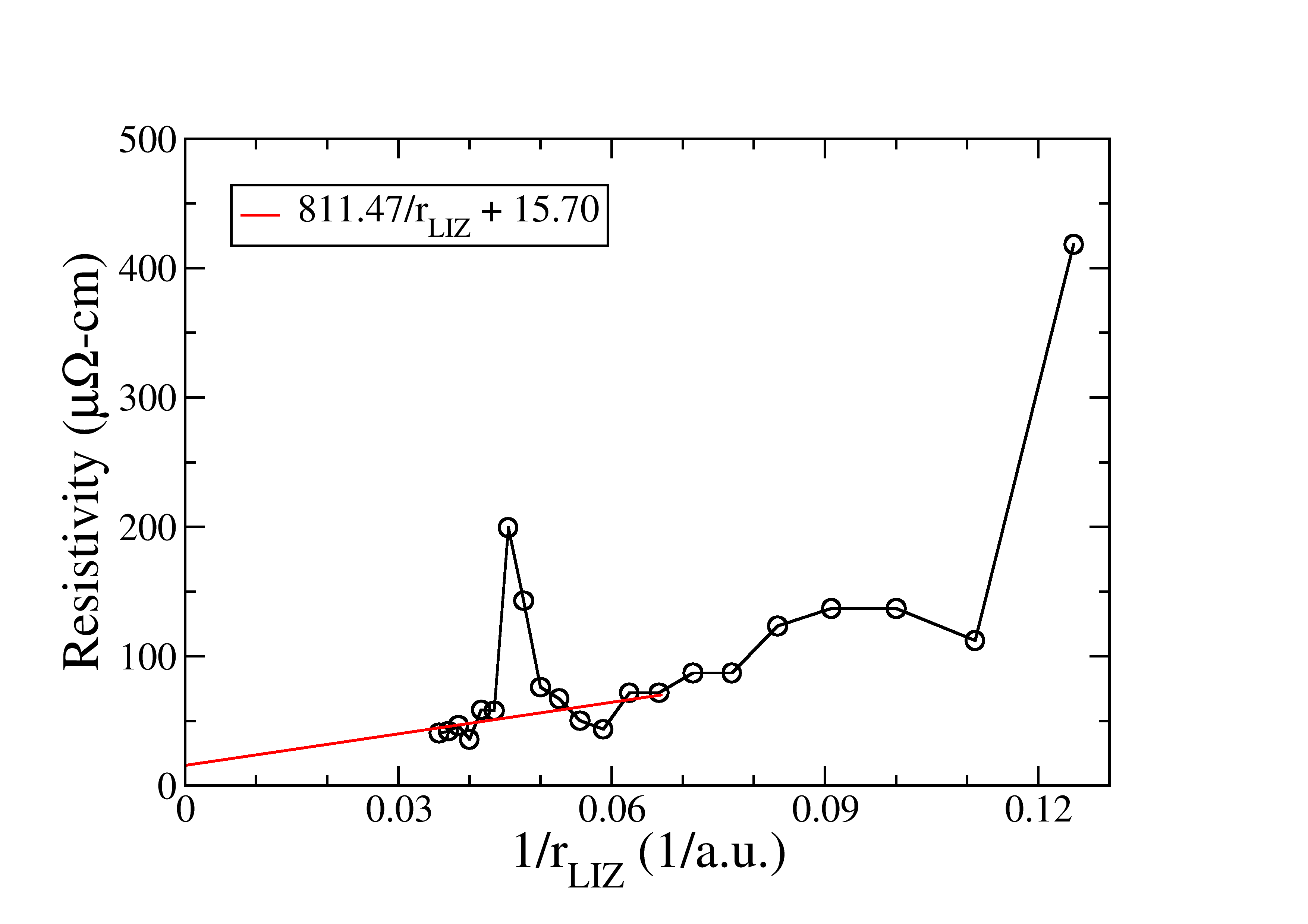}
       \caption{V}
       \label{fig:V-resistivity}
    \end{subfigure}
    \caption{The resistivity of pure elements as a function of the local interaction zone radius. A linear extrapolation is applied based on a small scattering approximation.}
    \label{fig:pure-element-resistivity}
\end{figure}

For a pure element at 0 K, the residual resistivity should be zero. We calculate LSMS resistivity of some pure elements as a basic test of the method.  Figure \ref{fig:pure-element-resistivity} shows the resistivity of pure Ag, Li, Al and V as a function of the inverse local interaction zone radius. In Ag and Li, the transport behavior is dominated by the valence $s$-electrons, while V has $d$-electron valence and Al has both $s$ and $p$-electron valence. Due to finite memory, extrapolation is necessary to obtain the resistivity at the asymptotic limit ($1/r_{\rm{LIZ}} \xrightarrow[]{}$ 0). Based on the heuristic expressions derived in the previous section, we apply a linear extrapolation to obtain the resistivity at infinite LIZ radius. For most of the pure elements, this matches the trends observed in Figure \ref{fig:pure-element-resistivity}. Vanadium shows anomalous behavior - there is a sharp peak which is not explained by the small scattering approximation. The extrapolated residual resistivity is small for all the cases. Since our heuristic models only produce the functional form, we are unable to determine why there is a slight underestimate for Ag but an overestimate for the other cases. More complex theoretical models are required to further understand these trends.
\subsection{Binary alloys}
\begin{figure}
   \centering
    \begin{subfigure}{0.49\linewidth}
       \centering
        \includegraphics[width=\linewidth]{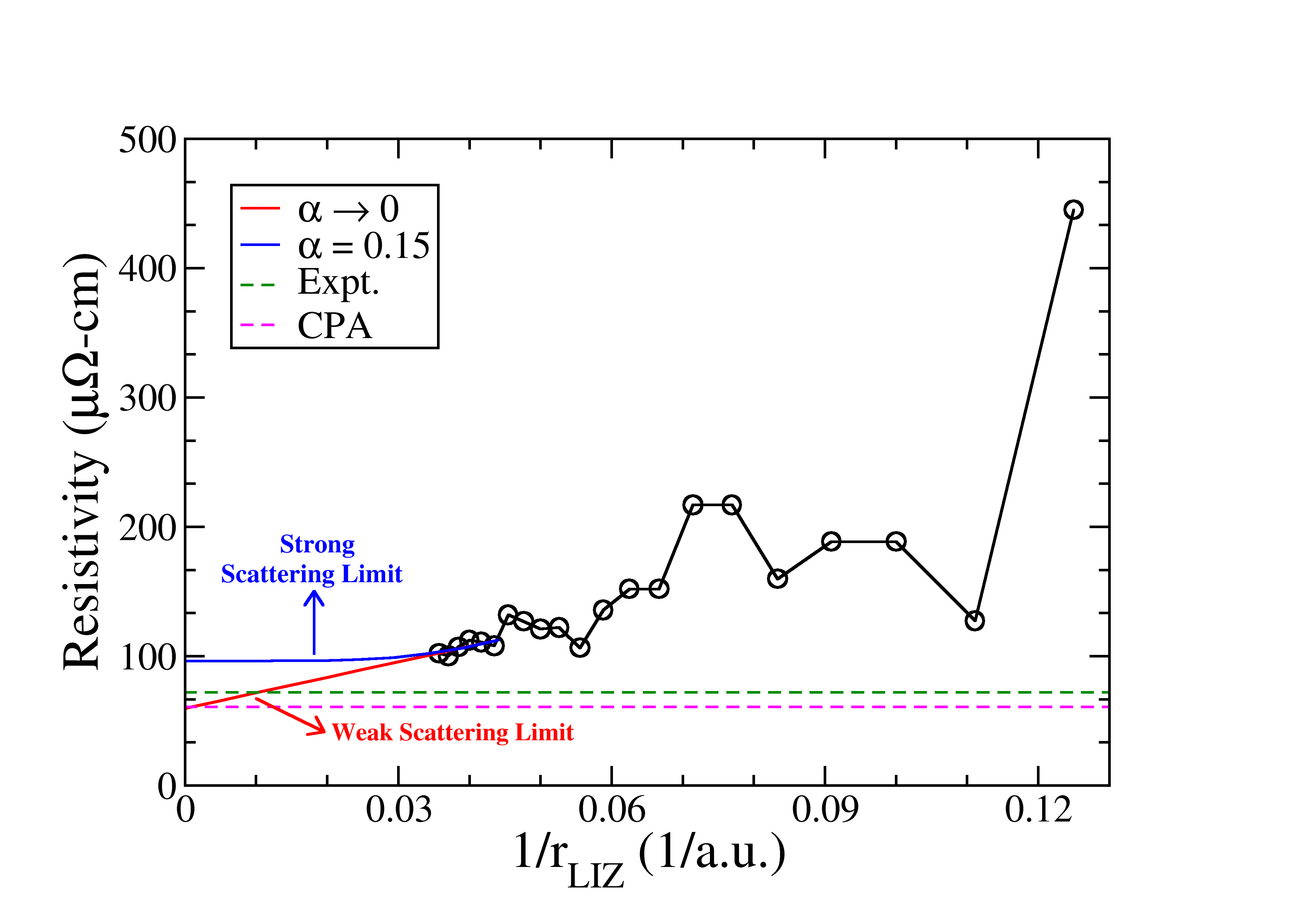}
        \caption{$x = 0.11$,
        $\rho^{\rm{lower}} = 1265/r_{\rm{LIZ}} + 67.07$, $\rho^{\rm{upper}} = 108.05/(1 - 3.41e^{-0.15r_{\rm{LIZ}}})$}
        \label{fig:AlV-0.11-LIZ}
    \end{subfigure}
    \begin{subfigure}[b]{0.49\linewidth}
        \includegraphics[width=\linewidth]{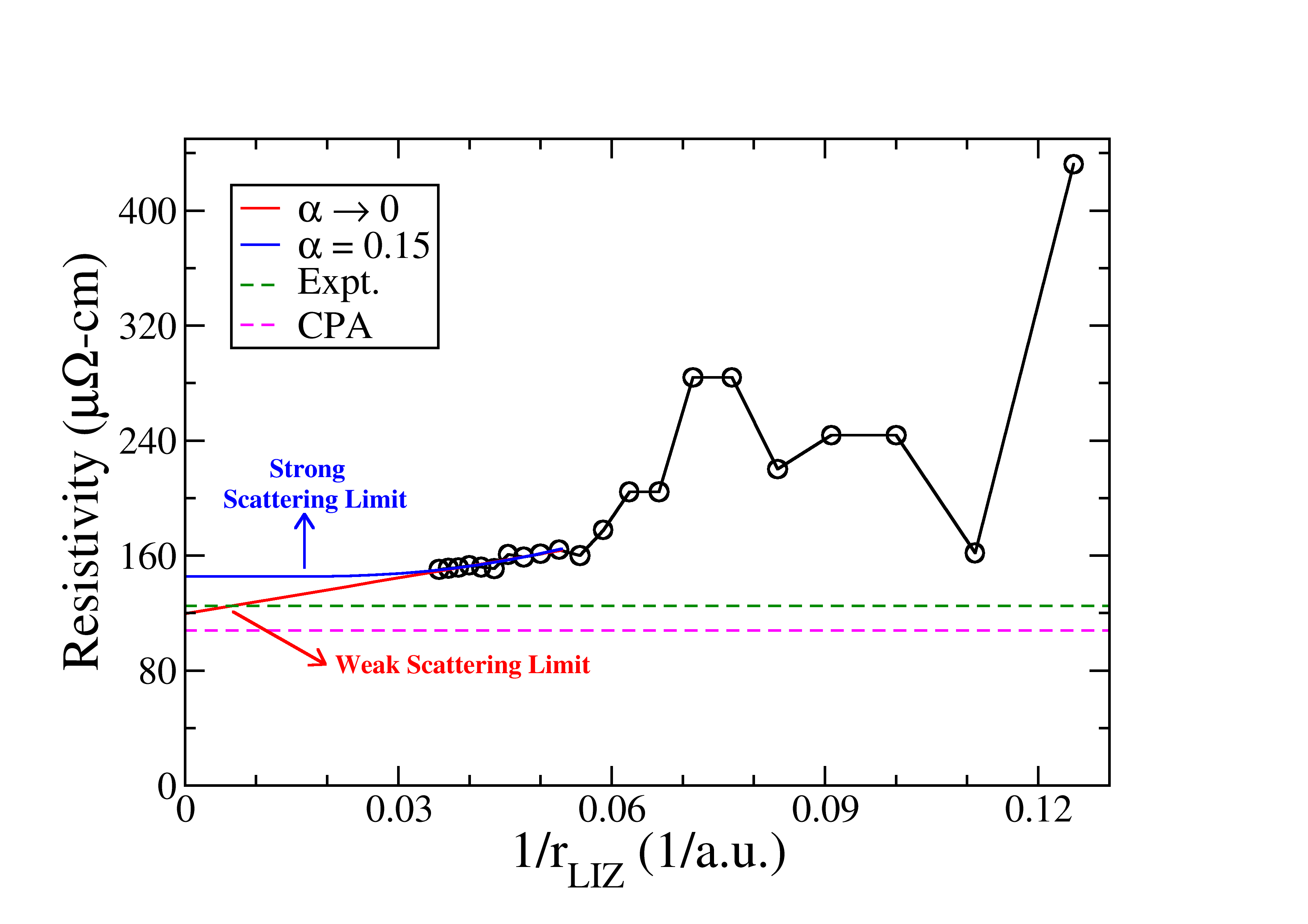}
        \caption{$x = 0.19$, $\rho^{\rm{lower}} = 1207/r_{\rm{LIZ}} + 118.8$, $\rho^{\rm{upper}} = 157.06/(1 - 2.45e^{-0.15r_{\rm{LIZ}}})$}
        \label{fig:AlV-0.19-LIZ}
    \end{subfigure}
    \begin{subfigure}[b]{0.49\linewidth}
        \includegraphics[width=\linewidth]{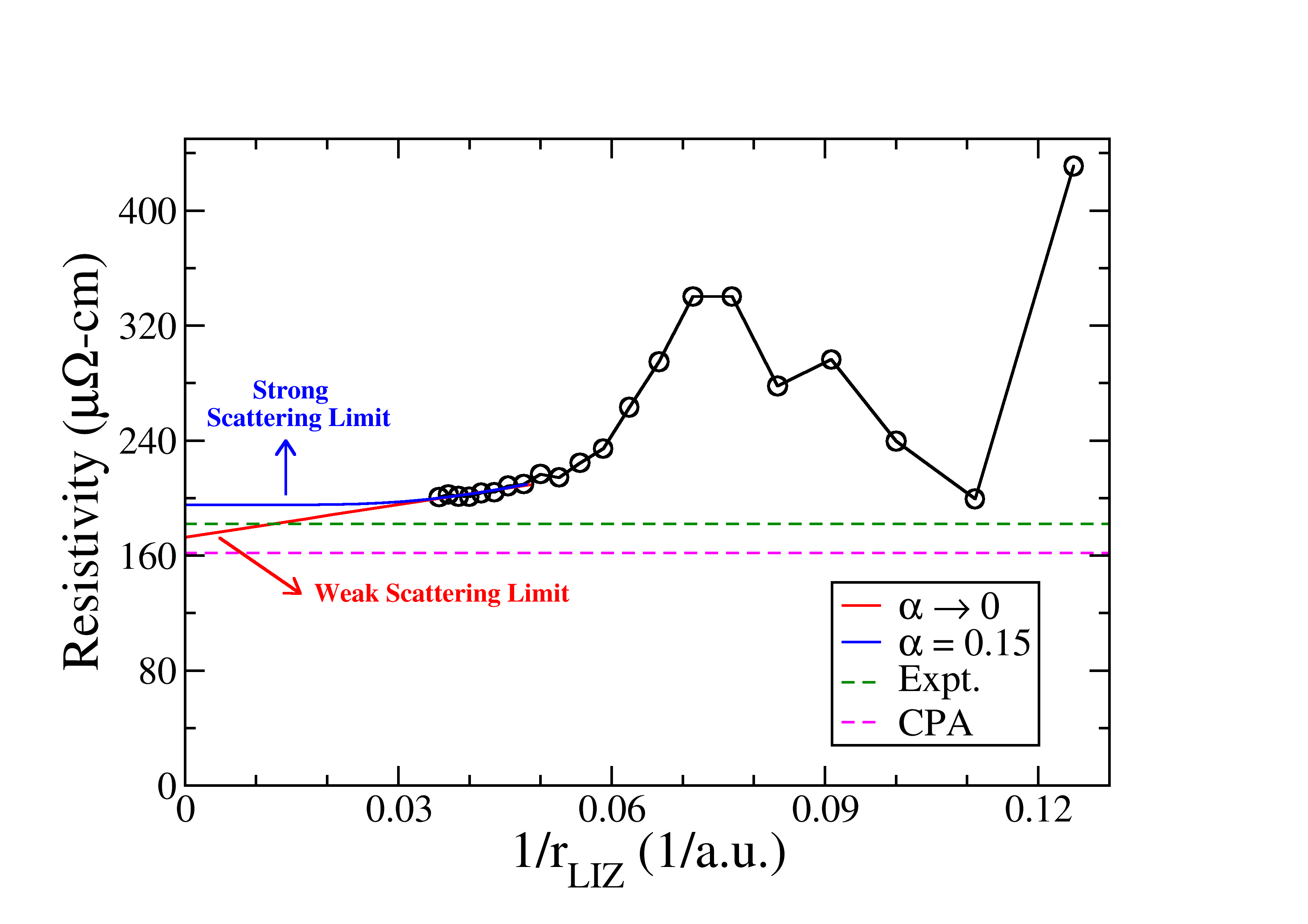}
        \caption{$x = 0.265$, $\rho^{\rm{lower}} = 1667/r_{\rm{LIZ}} + 173.9$, $\rho^{\rm{upper}} = 225.3/(1 - 2.74e^{-0.15r_{\rm{LIZ}}})$}
        \label{fig:AlV-0.265-LIZ}
    \end{subfigure}
    \begin{subfigure}[b]{0.49\linewidth}
        \includegraphics[width=\linewidth]{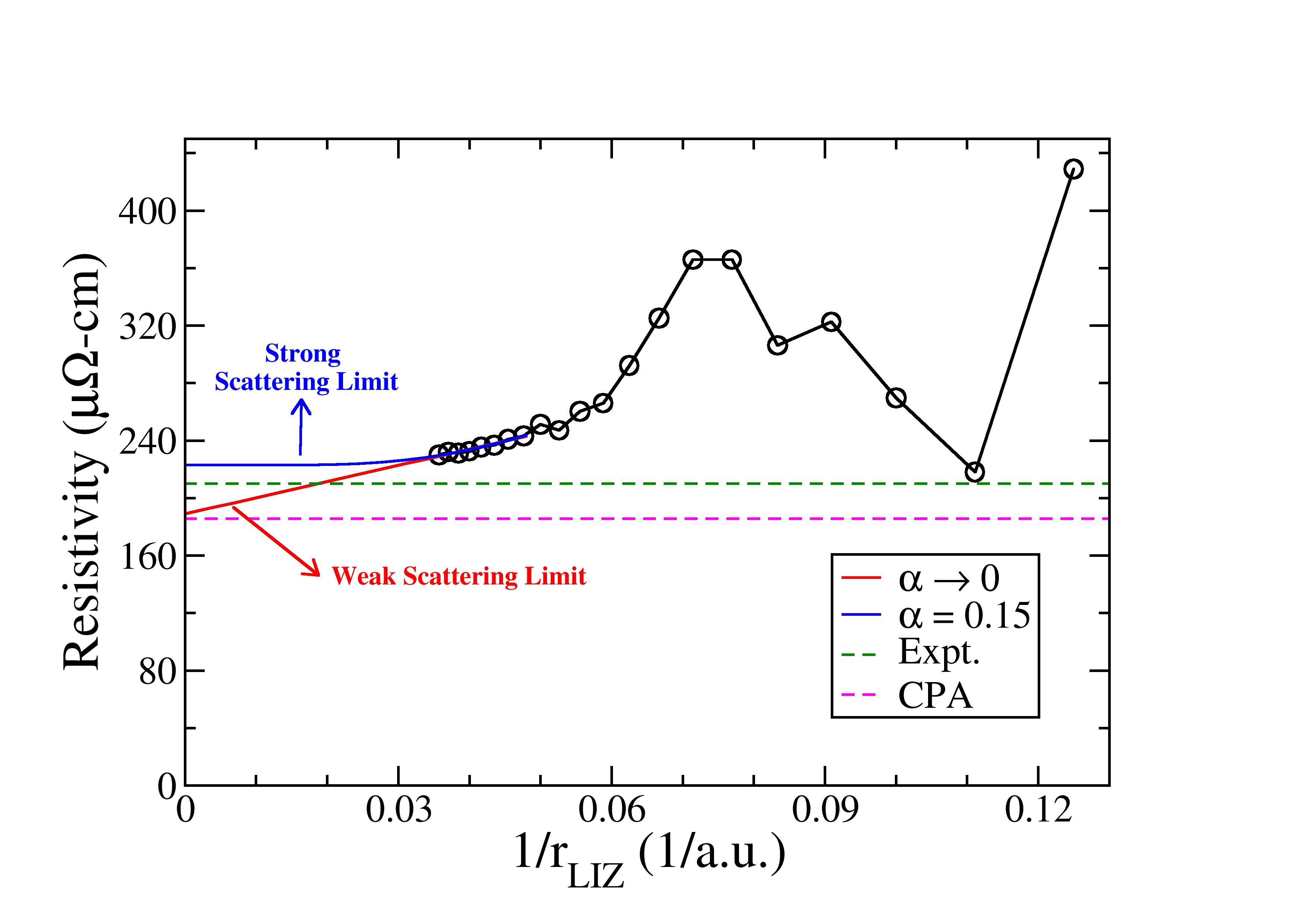}
        \caption{$x = 0.293$, $\rho^{\rm{lower}} = 1663/r_{\rm{LIZ}} + 214.8$, $\rho^{\rm{upper}} = 268.03/(1 - 1.83e^{-0.15r_{\rm{LIZ}}} )$}
        \label{fig:AlV-0.293-LIZ}
    \end{subfigure}
    \caption{The resistivity of Al$_x$V$_{1-x}$ as a function of the local interaction zone radius, denoted by the black line with circles. The experimental value is denoted by the dotted green line. Upper bound (blue line) and lower bound (red line) heuristic functional forms are fit and extrapolated to obtain the residual resistivity.}
\end{figure}
\begin{figure}[ht]\ContinuedFloat
    \begin{subfigure}[b]{0.49\linewidth}
        \includegraphics[width=\linewidth]{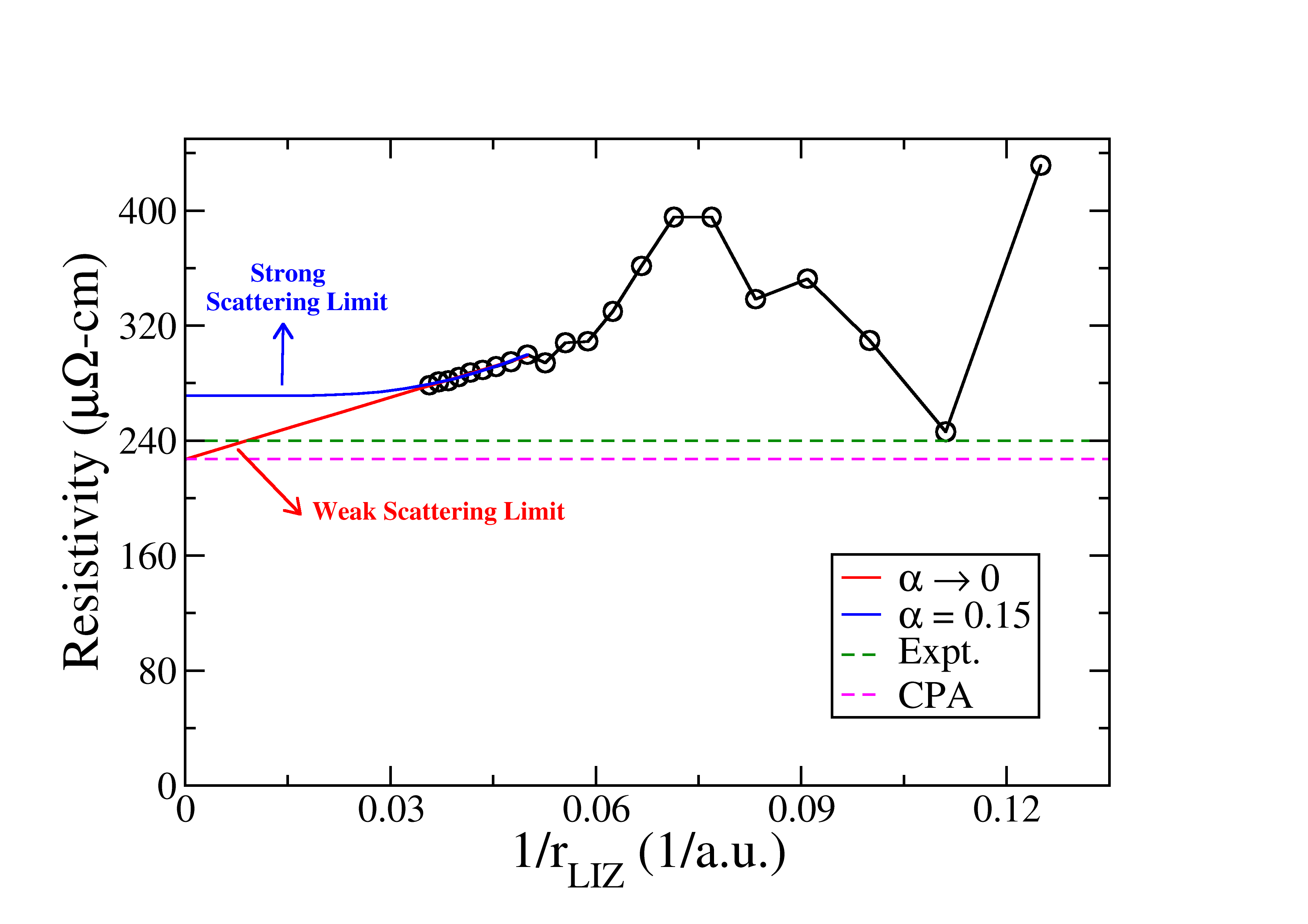}
        \caption{$x = 0.34$, $\rho^{\rm{lower}} = 1442/r_{\rm{LIZ}} + 277.8$, $\rho^{\rm{upper}} = 322.33/(1 - 1.63e^{-0.15r_{\rm{LIZ}}})$}
        \label{fig:AlV-0.34-LIZ}
    \end{subfigure}
    \caption{The resistivity of Al$_x$V$_{1-x}$ as a function of the local interaction zone radius, denoted by the black line with circles. The experimental value \cite{alv-expt} is denoted by the dotted green line. Upper bound (blue line) and lower bound (red line) heuristic functional forms are fit and extrapolated to obtain the residual resistivity. (contd.)}
    \label{fig:AlV-LIZ-plots}
\end{figure}
\begin{figure}
    \centering
    \includegraphics[width=0.6\linewidth]{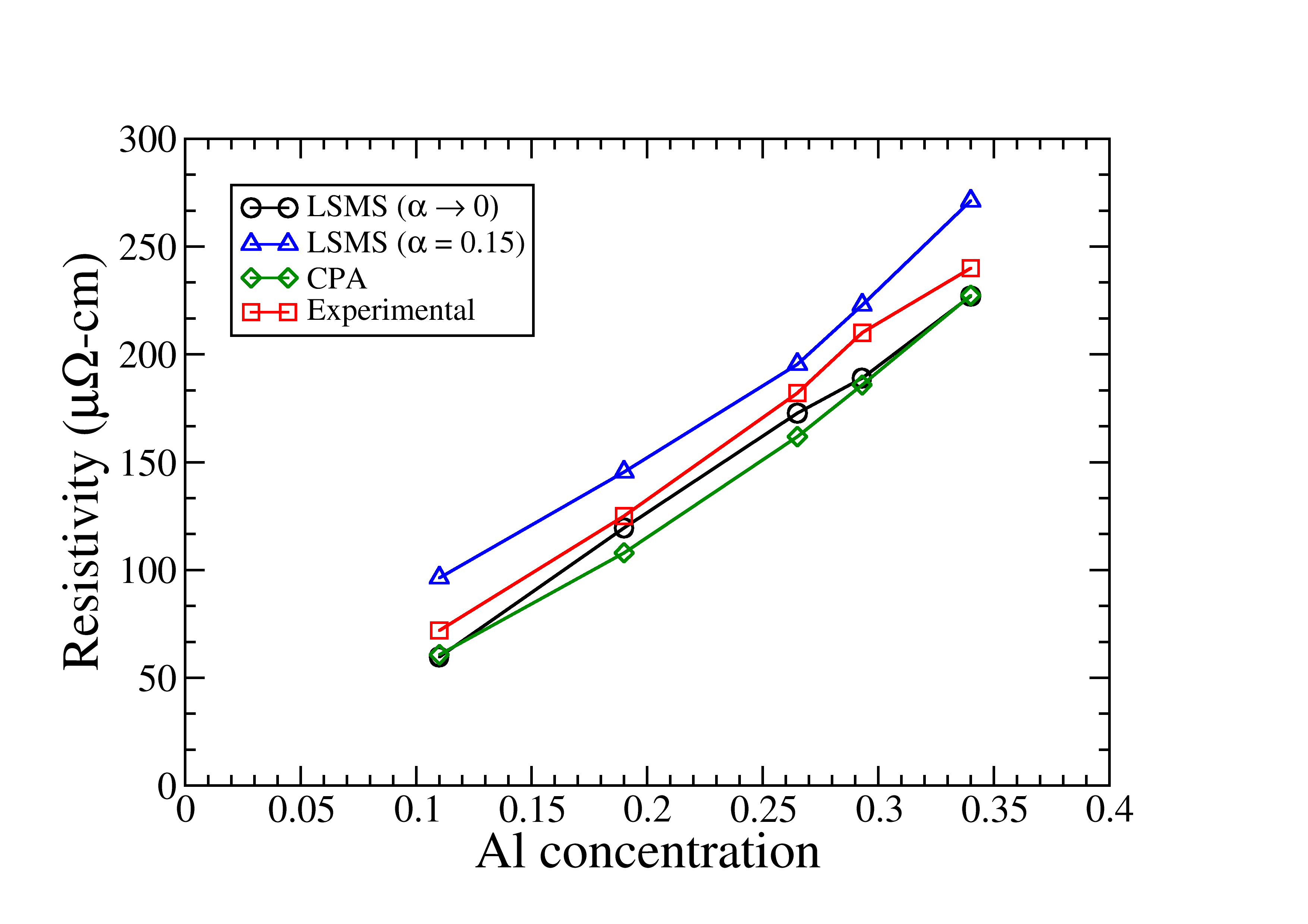}
    \caption{Residual resistivity of BCC Al$_x$V$_{1-x}$ as function of $x$. The black circles represent lower bound values calculated using the LSMS resistivity method, while the blue triangles represent the upper bound values. The green circles represent resistivity obtained using the coherent potential approximation (CPA), and the red squares represent the experimental values \cite{alv-expt}.}
    \label{fig:AlV-experimental-comparison}
\end{figure}

For random alloys, we expect the resistivity to decay exponentially to a non-zero value, in accordance with the heuristic derived in the previous section
\begin{equation}
    \rho_{\mu\nu} = \frac{b_0}{1 +  b_1 e^{-2\alpha r_{\rm{LIZ}}}}.
\end{equation}
Figure \ref{fig:AlV-LIZ-plots} shows the resistivity as a function of the inverse LIZ radius for BCC Al$_x$V$_{1-x}$ where $x=0.11,0.19,0.265,0.293$ and 0.34. We see a strong linear trend in the data, implying that the non-linear regime has not been reached. Hence, the $\alpha$ parameter cannot be determined by fitting the heuristic - any $\alpha$ value would yield an acceptable fit to the linear trend. As a result, we establish a range for $\alpha$, which we use to make upper and lower bound fits. We know that $\alpha$ is inversely proportional to the mean free path. In the strong scattering limit, the mean free path is expected to be close to the lattice spacing \cite{ioffe,ioffe-regel}. The strong scattering function should also fit the linear trend, and become non-linear almost immediately after the last data point. We choose $2\alpha_{\rm{max}} = 0.15\;\si{\angstrom}^{-1}$ (corresponding to a mean free path of 6-7 $\si{\angstrom}$), producing the upper bound function
\begin{equation}
\rho^{\rm{upper}}_{\mu\nu} = \frac{b_0}{1 + b_1e^{-0.15r_{\rm{LIZ}}}}.
\end{equation}
In the weak scattering limit, the mean free path is very large and $\alpha_{\rm{min}} \rightarrow 0$, resulting in a linear lower bound function
\begin{equation}
    \rho^{\rm{lower}}_{\mu\nu} \approx c_0\frac{1}{r_{\rm{LIZ}}} + c_1,
\end{equation}
which is similar to \ref{eq:linear-fit}, with $r_c \rightarrow 0$. The resistivity as a function of the LIZ radius can be bounded between the upper and lower bound functions
\begin{equation}
    \rho^{\rm{lower}}_{\mu\nu} \leq \rho_{\mu\nu} \leq \rho^{\rm{upper}}_{\mu\nu}
\end{equation}
Figure \ref{fig:AlV-LIZ-plots} shows these fits applied to the LSMS data. We also calculate first principles residual resistivity from the Korringa-Kohn-Rostoker Coherent Potential Approximation (KKR-CPA) method, an effective medium technique to model the random alloy. Figure \ref{fig:AlV-experimental-comparison} compares the extrapolated upper and lower bound LSMS resistivity to the experimental and CPA values. The experimental values \cite{alv-expt} lie between the two bounds for all the cases. The lower bound values compare very well with the experimental and CPA values. The upper bound values in some cases also compare well with the experiment. At all concentrations, the CPA values are an underestimate, which is a well known feature of resistivities obtained from KKR-CPA.

\begin{figure}
    \centering
    \includegraphics[width=0.6\linewidth]{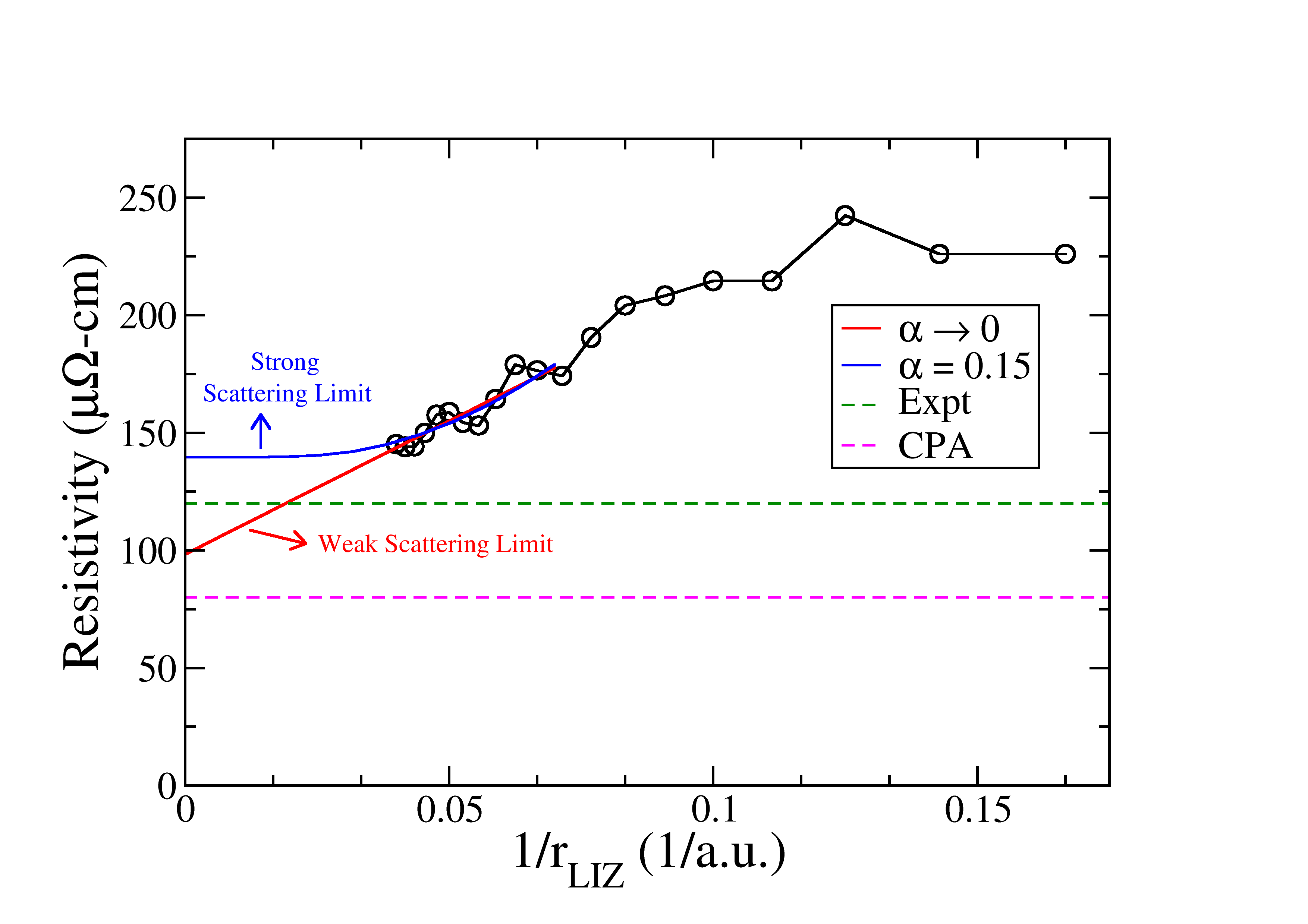}
    \caption{The resistivity of BCC Fe-9 wt\% Si as a function of the local interaction radius. The experimental value at room temperature is denoted by the dotted green line. Upper bound (blue line, $\rho = 139.67/(1 - 1.87e^{-0.15r_{\rm{LIZ}}})$) and lower bound (red line, $\rho = 1133.5/r_{\rm{LIZ}} + 98.33$)  fits are applied and extrapolated to obtain the residual resistivity.}
    \label{fig:FeSi-resistivity}
\end{figure}

We also calculate the resistivity of BCC Fe-9wt\% Si alloy (Figure \ref{fig:FeSi-resistivity}). A collinear spin polarized calculation was performed where the contribution of the spin up and spin down electrons to the conductivity was calculated separately and summed. The experimental value \cite{fesi-expt} lies in between the lower and upper bound estimates. The LSMS lower bound and KKR-CPA resistivity underestimate the resistivity, which is expected since the experiment was performed at room temperature.
\section{Conclusion}
In this work, we introduce the LSMS-Kubo-Greenwood technique, which combines the Kubo-Greenwood equation with first principles LSMS theory. This allows efficient calculation of electrical resistivity for very large systems. We implement this method in the high performance open source code $lsms$. Using a heuristic approach, we demonstrated that the conductivity should be linear and diverging for pure elements, and non-linear and convergent for alloys. The convergence is controlled by the parameter $\alpha$, which is dependent on the mean free path. In the small $\alpha$ limit, a convergent linear expression was obtained.  We test the method by applying it to pure elements, where a linear trend were observed and extrapolated to obtain very low resistivity. We also apply this method to binary random alloys Al$_x$V$_{1-x}$ and Fe-9wt\% Si. A linear trend was also observed in these systems which was fitted to an upper bound and lower bound functional form, determined from extremum values of $\alpha$. The experimental resistivities were found to lie between the two bounds, with the lower bound extrapolated values comparing well with experimental and KKR-CPA values. This demonstrates the validity of our approach.

There are areas of improvement in the theory and implementation of LSMS conductivity. While we are able to efficiently calculate resistivity for large unit cells and LIZ size of upto 1000 atoms, it should be possible to improve the code to allow for even larger LIZ sizes. The heuristic approach is very useful to understand the general trend and obtain upper and lower bounds. However a more precise functional form for the resistivity could be derived without introducing any ad-hoc parameters. Both these improvements are highly complicated and we are currently determining how they can be achieved.

The ability to deal with a large number of atoms opens the door for several novel and exciting applications. LSMS can be used to calculate the resistivity as a function of short range ordering in high entropy alloys. We can study the effect of stacking faults, dislocations and other defects on the electrical resistivity. We can also use LSMS to calculate the transport properties of non-crystalline systems like quasicrystals and metallic glasses. These applications are the subject of future work.

\begin{acknowledgements}
This work was supported by NSF under grant DMR-2103958. This research also used resources of the Oak Ridge Leadership Computing Facility, which is supported by the Office of Science of the U.S. Department of Energy under Contract No. DE-AC05-00OR22725. This research was supported in part by an appointment to the Oak Ridge National Laboratory GRO Program, sponsored by the U.S. Department of Energy and administered by the Oak Ridge Institute for Science and Education. The authors would like to thank D. Nicholson for providing helpful comments. 

\end{acknowledgements}

\end{document}